\documentstyle[proceedings]{crckapb}

\begin{opening}

\title{EXCHANGE CROSS SECTIONS FOR HARD BINARIES}

\author{DOUGLAS C. HEGGIE}
\institute{University of
Edinburgh, King's Buildings,\\ Edinburgh EH9 3JZ, U.K.}
\author{PIET HUT}
\institute{Institute for Advanced Study, Princeton, NJ 08540, U.S.A.}
\author{STEVEN L.W. MCMILLAN}
\institute{Drexel University, Philadelphia, PA 19104, U.S.A.}

\end{opening}
\begin{document}
\begin{abstract}
We present results on the exchange cross section for the
interaction between a hard binary and a field of single stars, for
arbitrary masses.  The  results are based partly on extensive numerical
scattering experiments, and partly on analytic estimates of the
mass-dependence of the cross section.  They can be used to estimate the
rate of exchange in an arbitrary mixture of masses, provided that the
binary is hard.
\end{abstract}

\section{Introduction}

In globular clusters, exchange
interactions between binary stars and single stars are a plausible 
mechanism for the formation of low-mass X-ray binaries.   The
rate of formation depends on the mass function of the stars and binaries
present.  Relatively little is known about the relevant cross sections
for unequal masses.  If they are computed by three-body scattering
experiments, a fresh set of experiments must be performed for each mass
function.  Here we summarise a forthcoming  paper (Heggie {\sl et al.} 1995) in which we  present
approximate cross sections which are valid  for
arbitrary masses if the binary is hard.  Then the rate of formation is
given by a simple integration over the mass function.

\section{Analytical Estimates}

For comparable masses the hard binary exchange cross section is easily
estimated by correcting the geometrical area of the binary for
gravitational focusing.  The result is
$
\Sigma \sim GM_{123}a/ V^2,
$
where $M_{123}$ is the total mass of the three participating stars, $a$
is the initial semi-major axis of the binary, and $V$ is the speed of
the intruder relative to the barycentre of the binary, while they are
still far apart.

The next step is to investigate how this formula is modified in various
extreme mass regimes.  For example, suppose the mass of the intruder,
$m_3$, much exceeds that total mass of the binary, $M_{12}$.  Then the
binary can be unbound by a tidal interaction, and the corresponding
cross section is quite readily estimated.  
Surprisingly, all these estimates (in the various extreme mass regimes)
can be summarised in a single formula, which is given by the coefficient
of the exponential in the equation below. 

\section{Inclusion of Numerical Data}

We have used a scattering package within the Starlab
environment (Hut \& McMillan 1995) to compute accurate cross
sections for exchange.  Initial conditions were chosen
appropriate to hard binaries with a Jeans eccentricity
distribution, for a wide variety of masses.

The numerical results have been combined with our analytic estimate in
the following semi-numerical fitting formula.  Let $x =
m_1/M_{12}$,  $y = m_3/M_{123}$ and $M_{ij} = m_i + m_j$,
where $m_1$ is the ejected component.  Then
\begin{eqnarray*}
\Sigma &\simeq& {GM_{123}a\over V^2}{m_3^{7/2}M_{23}^{1/6}\over
M_{12}^{1/3} M_{13}^{5/2}M_{123}^{5/6}}
\exp (3.70 + 7.49x -1.89 y 
- 15.49x^2 -\\
&-&2.93xy - 2.92y^2 + 
3.07x^3 + 13.15x^2y
-5.23xy^2 + 3.12y^3).\\
\end{eqnarray*}

This formula fits 75\% of the measurements to better than 20\%; larger
discrepancies are restricted to mass ratios where the cross sections are
probably too small to be of importance in applications.  We have also
compared our formula with results by other authors, and again the
agreement is satisfactory in general.  In addition, these comparisons
suggest that the above formula is applicable also to circular binaries,
except in some parameter ranges where the cross section is very small
anyway.

\bigskip
\noindent
{\bf References}
\smallskip


Heggie D.C., Hut P., McMillan S.L.W., 1995, preprint

Hut P., McMillan S.L.W., 1995, preprint


\end{document}